\documentclass[journal]{IEEEtran} 
\usepackage{amsmath, amssymb, amsfonts, amsbsy, mathrsfs, bm , bbm}
\usepackage{cite, enumerate}
\usepackage[table]{xcolor}
\usepackage[normalem]{ulem}
\usepackage{dsfont} 
\usepackage{algorithm}
\usepackage{algpseudocode}
\usepackage{subfigure}
\usepackage{graphicx}
\usepackage{epstopdf}
\usepackage{balance}

\newtheorem{remark}{Remark}
\newtheorem{assumption}{Assumption}

\newtheorem{lem}{Lemma}

\def\R{{\mathbb{R}}}

\def\Ee{{\mathbb{E}}}

\newcommand{\E}[1]{\mathbb{E}\left[#1 \right]}
\newcommand{\col}[1]{\mathrm{col}\left\{#1 \right\}}
\newcommand{\Diag}[1]{\mathrm{diag}\left\{#1 \right\}}
\newcommand{\vect}[1]{\mathrm{vec}\left\{#1 \right\}}

\def\bx{{\mathbf{x}}}
\def\by{{\mathbf{y}}}
\def\bn{{\mathbf{n}}}
\def\bv{{\mathbf{v}}}
\def\bpsi{{\pmb{\psi}}}
\def\b1{{\mathds{1}}}

\def\bH{{\mathbf{H}}}
\def\bF{{\mathbf{F}}}
\def\bG{{\mathbf{G}}}
\def\bR{{\mathbf{R}}}
\def\bQ{{\mathbf{Q}}}

\def\bI{{\mathbf{I}}}
\def\bP{{\mathbf{P}}}
\def\bS{{\mathbf{S}}}
\def\bT{{\mathbf{T}}}
\def\bB{{\mathbf{B}}}

\def\bmS{{\boldsymbol{\mathcal{S}}}}
\def\bmB{{\boldsymbol{\mathcal{B}}}}
\def\bmX{{\boldsymbol{\mathcal{X}}}}
\def\bmP{{\boldsymbol{\mathcal{P}}}}
\def\bmH{{\boldsymbol{\mathcal{H}}}}
\def\bmF{{\boldsymbol{\mathcal{F}}}}
\def\bmG{{\boldsymbol{\mathcal{G}}}}
\def\bmD{{\boldsymbol{\mathcal{D}}}}
\def\bmO{{\boldsymbol{\mathcal{O}}}}
\def\bmL{{\boldsymbol{\mathcal{L}}}}
\def\bmQ{{\boldsymbol{\mathcal{Q}}}}
\def\bmK{{\boldsymbol{\mathcal{K}}}}
\def\bmfB{{\boldsymbol{\mathfrak{B}}}}
\def\bmSig{{\boldsymbol{{\Sigma}}}}
\def\bmGa{{\boldsymbol{{\Gamma}}}}
\def\bmPsi{{\boldsymbol{{\Psi}}}}

\def\MSD{{\mathrm{MSD}}}

 \usepackage{graphicx}

\begin{document}
%
\title{Partial Diffusion Kalman Filtering}

\author{Vahid Vahidpour, Amir~Rastegarnia, Azam Khalili, Wael Bazzi, and Saeid Sanei \IEEEmembership{Senior Member,~IEEE} 
\thanks{
V. Vahidpour, A. Rastegarnia and A. Khalili are with the Department of Electrical Engineering, Malayer University, Malayer, Iran (e-mail: v.vahidpur@ieee.rg, khalili@malayeru.ac.ir, rastegar@alayeru.ac.ir).

W. M. Bazzi is with the Department of Electrical Engineering, American University in Dubai, Dubai, United Arab Emirates, (email: wbazzi@and.edu)

S. Sanei is with the Department of Computer Science, University of Surrey, Surrey GU2 7XH, UK (email: s.sanei@surrey.ac.uk)

}
}


\IEEEpubid{0000--0000/00\$00.00~\copyright~2017 IEEE}

\maketitle

\begin{abstract}
In conventional distributed Kalman filtering, employing diffusion strategies, each node transmits its state estimate to all its direct neighbors in each iteration. In this paper we propose a partial diffusion Kalman filter (PDKF) for state estimation of linear dynamic systems. In the PDKF algorithm every node (agent) is allowed to share only a subset of its intermediate estimate vectors at each iteration among its neighbors, which reduces the amount of internode communications. We study the stability of the PDKF algorithm where our analysis reveals that the algorithm is stable and convergent in both mean and mean-square senses. We also investigate the steady-state mean-square deviation (MSD) of the PDKF algorithm and derive a closed-form expression that describes how the algorithm performs at the steady-state. Experimental results validate the effectiveness of PDKF algorithm and demonstrate that the proposed algorithm provides a trade-off between communication cost and estimation performance that is extremely profitable.
\end{abstract}


%

\IEEEpeerreviewmaketitle

\section{Introduction}
\label{sec1}
We consider the problem of distributed Kalman filtering over a set of interconnected nodes called adaptive network (AN). These nodes are closely bound up with each other and are able to cooperatively perform decentralized data processing and optimization through locally information exchange. We assume that the system of interest is described as a linear state-space model, and that each individual node in the network assembles measurements that are linearly linked to the unobserved state vector. Estimating the state of the system is the objective for every node. As opposed to the centralized estimation, the  distributed estimation is more flexible for topology changes and robust to node/link failures. 

In previous studies, some decentralized Kalman filtering algorithms such as parallel information filter \cite{speyer1979computation}, where a centralized control problem is formulated based on the centralized Kalman filtering algorithm, distributed information filter \cite{rao1991fully}, distributed Kalman filter with consensus filter \cite{spanos2005approximate,olfati2005distributed,olfati2007distributed}, distributed Kalman filter with weighted averaging \cite{estrin2001instrumenting,alriksson2006distributed}, and distributed Kalman filter with diffusion strategies \cite{Cattivelli2010,cattivelli2008diffusion} have been proposed in the context of network of nodes. Here, we focus on diffusion version of the Kalman filter (DKF) proposed in \cite{Cattivelli2010}, where the nodes are tasked with estimating some parameters of interest, describing the state of the system, from noisy measurements through a diffusion cooperative protocol. 

Cooperation structure in diffusion-based adaptive networks makes them scalable and more robust to link/node failures. In diffusion strategies, nodes communicate with all their immediate neighbors to share their intermediate estimates with each other \cite{Cattivelli2010a}. However, due to limited power and bandwidth resources for communication among the nodes over a practical sensor network, the most expensive part of realizing a cooperative task is data transmission through radio links. Therefore, lowering the amount of information exchange among the eighbor nodes, while keeping the benefits of cooperation, is of practical importance. Generally speaking, although the benefits of diffusion strategies are achieved by increasing internode communications, they are compromised by communication cost. 

There have been several efforts to reduce the communication cost without any significant degradation of the estimation and compromising the cooperation benefits in diffusion algorithms, such as reducing the dimension of the estimate \cite{Sayin2013,Sayin2014,Chouvardas2013}, selecting a subset of the entries of the estimates \cite{Arablouei2014a,Arablouei2014,vahid15,vahid17}, set-membership filtering \cite{deller2002set,Gollamudi1998,deller1993least} or partial updating \cite{Dogancay2008} have been reported in \cite{werner2010energy,malipatil2009smf,Werner2009,Werner2008,werner2010time}. All these correspondences aim at  reducing the internode communication in diffusion strategies, notably diffusion least mean-square (LMS). In \cite{khan2008distributing}, the authors addressed the problem of distributed Kalman filter and propose an efficient algorithm for large-scale systems. A distributed Kalman filtering with low-cost communications has been reported in \cite{ribeiro2006soi}. The algorithm relies on the average consensus and employs the sign of innovations to reduce the communication cost to a single bit per observation. Kalman filters with reduced order models have been studied, in e.g., \cite{berg1991model,mutambara1998decentralized}, to address the computation burden posed by implementing nth order models. In these works, the reduced models are decoupled, which is sub-optimal as the important coupling among the system variables is ignored. 
\IEEEpubidadjcol 
Partitioned Update Kalman Filter (PUKF) that updates the state using multidimensional measurements in parts is discussed in \cite{raitoharju2015partitioned}. PUKF evaluates the nonlinearity of the measurement fusion within a Gaussian prior by comparing the effect of the second order term on the Gaussian measurement noise. Among these methods, we focus on partial-diffusion based algorithms \cite{Arablouei2014a,Arablouei2014} in which the nodes only selected and diffused a subset of their estimate vector entries through the network. 

In this paper, we employ partial-diffusion strategy in Kalman filtering to collectively estimate the state of a linear system, in which each node exchange and diffuse a part of its estimate state vector only with its direct neighbors at each time update. Inspired by \cite{Arablouei2014a}, we name the proposed algorithm partial-diffusion Kalman filtering (PDKF) algorithm. The PDKF algorithm lowers the total amount of internode transmissions in the network related to the diffusion Kalman filtering (DKF) algorithm, where the nodes always receive the intermediate estimates of all their neighbors, with limited degradation in performance. To select a subset of estimate vector for broadcasting at each iteration, we consider two similar schemes, sequential and stochastic, proposed in \cite{Arablouei2014a}. 

The main contributions of this paper can be summarized as follows:
\begin{enumerate}[(i)]
	\item We employ partial-diffusion algorithm in the Kalman filtering algorithm. More specifically, we adopt a similar approach to that proposed in \cite{Arablouei2014a,Arablouei2014} to build up the PDKF algorithm. It should be noted that since our objective is to minimize the internode communication, nodes exchange their intermediate estimates with their neighbors only and do not exchange the local data;
	\item Using the energy conservation argument \cite{sayed2011adaptive,yousef2001unified,rupp1996time} we analyze the stability of algorithms in mean and mean square sense under certain statistical conditions;
	\item Stability conditions for PDKF algorithm are derived and interpreted;
	\item We derived closed-form expression for mean-square derivation (MSD) to explain the steady-state performance of PDKF algorithm;
	\item We illustrate the comparable convergence performance of PDKF algorithm in different numerical examples.
\end{enumerate}
 Using the PDLMS algorithm, acceptable estimation performance is achieved while the utilization of communicated resource is kept low. The main aim of this correspondence is that in comparison to DKF when part of the intermediate estimate is received by the neighbors the total amount of internode communication is reduced. This fact comes at the expense of slight degradation in performance. The more entries are communicated at each iteration, the more weight estimates are interred into the consultation phase. So, the PDKF algorithm enables an effective trade-off between communication cost and estimation performance. The effectiveness of the proposed PDKF algorithm as well as the accuracy of the theoretical derivations are validated with simulation results.
 
The rest of this correspondence is organized as follows. In section II.A, we briefly introduce the data model, the diffusion KF algorithm, and the local estimator of estate vector, to offer preliminary knowledge. In section II.B, the partial-diffusion KF algorithm is formulated. The performance analyses are examined in section IV. In section V, numerical simulations are presented to illustrate the effectiveness and steady-state behavior of the proposed algorithm. Finally, conclusions are drawn in section V.


\textit{Notation}: We adopt small boldface letters for vectors and bold capital letters for matrices. The symbol $^*$ denotes conjugation for scalars and Hermitian transpose for matrices. The notation $\Diag{\cdot}$ is used in two ways: $\mathbf{X}=\Diag{\mathbf{x}}$ is a diagonal matrix whose entries are those of the vector $\mathbf{x}$, and $\mathbf{x}=\Diag{\mathbf{X}}$ is a vector containing the main diagonal of $\mathbf{X}$. The exact meaning of this notation will be clear from the context. If $\boldsymbol{\Sigma}$ is a matrix, we use the notation $\|\mathbf{x}\|_{\boldsymbol{\Sigma}}^2=\mathbf{x}^* \boldsymbol{\Sigma} \mathbf{x}$ for the weighted square norm of $\mathbf{x}$. Finally, we use $\otimes$ and $\mathds{1}$ to denote the Kronecker product a column vector with unity entries, respectively. 


\section{Algorithm Description}\label{sec2}
\subsection{Diffusion Kalman Filter Algorithm}
\begin{figure} [t]
\centering 
\includegraphics [width=6cm]{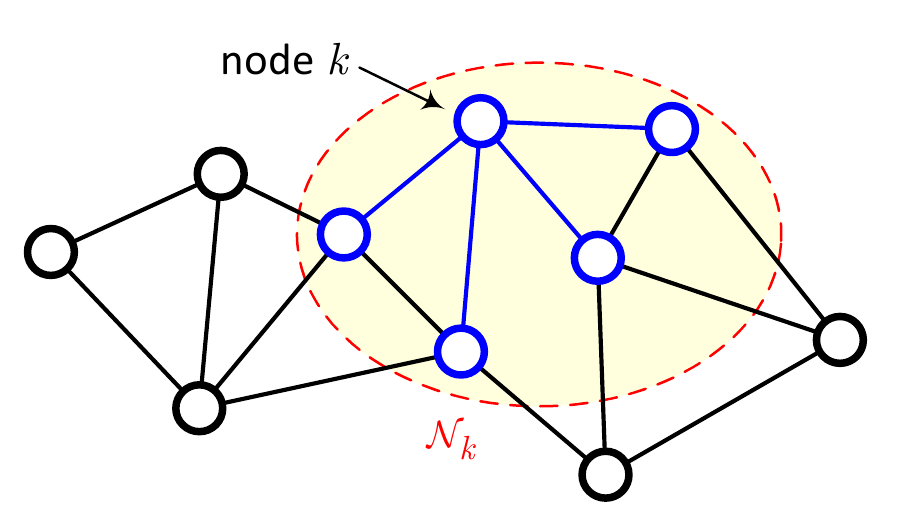} 
\centering \caption{At each time instant $i$, node $k$ collects a measurement $\by_{k,i}$.}
\label{fig:1}
\end{figure}

Consider a network of $N$ nodes scattered in geographical space. Two nodes are neighbors if they can exchange information with each other. The neighborhood of node $k$ is denoted by ${\cal N}_k$ (notice that $k\in \mathcal{N}_k$ (See Fig. \ref{fig:1}). At time instant $i$, each node $k$ collects an observation (or measurement) $\by_{k,i}\in\R^{P}$ of the true state $\bx_i\in\R^{M}$ of a linear dynamic process (system) as
\begin{equation}
\by_{k,i}=\bH_{k,i}\bx_i+\bv_{k,i}
\label{eq:1}
\end{equation}
where $\bH_{k,i}\in\R^{P\times M}$ denotes the local observation matrix and $\bv_{k,i}$ is the observation noise vector. The state vector $\bx_i$ evolves according to
\begin{equation}
\bx_{i+1}=\bF_i \bx_i+\bG_i \bn_i
\label{eq:1b}
\end{equation}
where $\bF_i\in\R^{M\times M}$ and $\bG_i\in\R^{M\times M}$ and $\bn_i\in\R^{M}$ denote the model matrix, the state noise matrix and the state noise vector respectively. 
For the state-space model described by \eqref{eq:1} and \eqref{eq:1b} it is customary to make the following assumptions:
\begin{assumption}\label{asp:1}\
\begin{enumerate}[(i)]
\item The measurement noises $\bv_{k,i}$ are zero-mean and spatially and temporally uncorrelated with covariance matrices given by $\mathbb{E}[\bv_{k,i}\bv_{l,j}^*]=\delta_{ij}\delta_{kl} \bR_{k,i} $.
\item The state noise vectors $\bn_i$ are zero-mean, and temporally uncorrelated with covariance matrices $\mathbb{E}[\bn_{i}\bn_{j}^*]=\delta_{ij} \bQ_i$. 
\item The noise signals $\bv_{k,i}$ and $\bn_i$ are spatially and temporally uncorrelated.
\item The initial state $\bx_0$ is zero-mean with covariance matrix $\E{\bx_0\bx^{*}_{0}}=\boldsymbol{\Pi}_0>0$ and uncorrelated with noise signals $\bv_{k,i}$ and $\bn_i$ for all $i$ and $k$.
\end{enumerate}
\end{assumption}
For every node in the network the objective is to estimate the state vector $\bx_i$ by collaborating with other nodes. As we mentioned earlier, the DKF algorithm is an effective tool for performing network-wide distributed Kalman filtering problem.  Before proceeding further, we define $\hat\bx_{k,i|j}$ as the local estimator of $\bx_i$ that node $k$ computes at time $i$ based on local observations and information up to and including time $j$. We further  use $\tilde{\bx}_{k,i|j}=\bx_i-\hat{\bx}_{k,i|j}$ to denote the estimation error at node $k$ and  $\bP_{k,i|j}$ to denote the covariance matrix of $\tilde{\bx}_{k,i|j}$. Then, the DKF algorithm in its time-and-measurement update form is given in Algorithm \ref{alg-1}. The algorithm starts with $\hat\bx_{k,0|-1}=\mathbf{0}$ and $\bP_{k,0|-1}=\boldsymbol{\Pi}_0$, where $\bP_{k,0|-1}\in\R^{M\times M}$. The symbol $\leftarrow$ denotes a sequential assignment.
 As can be seen, the algorithm consists of two steps, i.e. the \emph{incremental update} and  \emph{diffusion update}. In the incremental update step, first, the nodes exchange local data $\{y_{k,i},\bH_{k,i},\bR_{k,i}\}$ with their neighbors. Then, each node performs KF with available data to obtain the intermediate estimates ${\bpsi _{k,i}}$ as follows:
\begin{align} \label{eq:2}
&{\bpsi_{k,i}}  \leftarrow {\hat\bx_{k,i|i-1}}  	\nonumber  \\
&{\bP_{k,i}}  \leftarrow {\bP_{k,i|i-1}}  	\nonumber  \\  
&\mathrm{for}\ {\ell \in \mathcal{N}_k}   \nonumber  \\  
&\hspace{1cm}{\bR_{e,i}}  \leftarrow {\bR_{\ell,i}} + {\bH_{\ell,i}}{\bP_{k,i|i}}\bH_{\ell,i}^* \nonumber \\
&\hspace{1cm}{\bpsi _{k,i}}  \leftarrow {\bpsi_{k,i}} + {\bP_{k,i|i}}\bH_{\ell,i}^*{\bR}_{e,i}^{-1}[{\by_{\ell,i}}-{\bH_{\ell,i}}{\bpsi_{k,i}}]   \nonumber \\
& \hspace{1cm}{\bP_{k,i}}  \leftarrow {\bP_{k,i|i}}-{\bP_{k,i|i}}\bH_{\ell,i}^*\bR_{e,i}^{-1}{\bH_{\ell,i}}{\bP_{k,i|i}} \nonumber  \\ 
& \mathrm{end}   
\end{align}
In the diffusion step, the nodes share intermediate estimates ${\bpsi _{k,i}}$ and then compute a convex combination of intermediate estimates to obtain the local estimate ${\hat \bx_{k,i|i}}$ as:
\begin{equation}\label{eq:3}
{\hat \bx_{k,i|i}} \leftarrow \sum_{l\in{\cal N}_{k}}c_{lk}\bpsi_{l,i}
\end{equation}
It is noteworthy that, in \eqref{eq:3} the scalars $\{c_{lk}\}$ are nonnegative coefficients satisfying
\begin{equation}
\sum\limits_{l = 1}^N {c_{lk}} = 1,\quad  {c_{lk}} = 0\  \textrm{if}\ l \notin {\cal N}_k,\ \forall l,k
\label{eq:4}
\end{equation}

\begin{algorithm}[!t]
  \caption{Diffusion Kalman filter \cite{Cattivelli2010}}
  \begin{algorithmic}
	\State \textbf{Initialization:} $\hat\bx_{k,0|-1}=0$ and $\bP_{k,0|-1}=\boldsymbol{\Pi}_0$
		     \State For every time instant $i$, every node $k$ computes
			\State 
       \State \textbf{Step1:} Incremental Update
        \State ${\bpsi_{k,i}} \leftarrow {\hat\bx_{k,i|i-1}}$
		    \State ${\bP_{k,i}} \leftarrow {\bP_{k,i|i-1}}$  
				\For {$\ell \in \mathcal{N}_k$}
				\begin{align}
					{\bR_{e,i}} & \leftarrow {\bR_{\ell,i}} + {\bH_{\ell,i}}{\bP_{k,i|i}}\bH_{\ell,i}^* \nonumber \\
					{\bpsi _{k,i}} & \leftarrow {\bpsi_{k,i}} + {\bP_{k,i|i}}\bH_{\ell,i}^*{\bR}_{e,i}^{-1}[{\by_{\ell,i}}-{\bH_{\ell,i}}{\bpsi_{k,i}}]   \nonumber \\
          {\bP_{k,i}} & \leftarrow {\bP_{k,i|i}}-{\bP_{k,i|i}}\bH_{\ell,i}^*\bR_{e,i}^{-1}{\bH_{\ell,i}}{\bP_{k,i|i}}
					\nonumber
					\end{align}
					\EndFor
			 \State 
       \State \textbf{Step2:} Diffusion Update
			\begin{align}
{\hat \bx_{k,i|i}} &\leftarrow \sum_{l\in{\cal N}_{k}}c_{lk}\bpsi_{l,i}
  \nonumber  \\
{\bP_{k,i|i}} & \leftarrow {\bP_{k,i}} \nonumber \\
{\hat\bx_{k,i+1|i}}&=\bF_{i}{\hat\bx_{k,i|i}} \nonumber \\
{\bP_{k,i+1|i}} &= \bF_{i}{\bP_{k,i|i}}\bF_{i}^*+\bG_{i}\bQ_{i}\bG_{i}^*   \nonumber 
\end{align}
	\end{algorithmic}
  \label{alg-1}
  \end{algorithm}

\subsection{PDKF Algorithm Derivation}
In this correspondence, we adopt a similar approach proposed in \cite{Arablouei2014a} to build up our PDKF algorithm. 
In the DKF given in Algorithm \ref{alg-1} the nodes exchange local data $\{y_{k,i},\bH_{k,i},\bR_{k,i}\}$ to calculate the intermediate estimates ${\bpsi _{k,i}}$. Clearly, successful implementation of such strategy requires considerable communication resources. Thus, in order to reduce the communication complexity, our proposed PDK algorithm relies on the modified version of DKF given in Algorithm \ref{alg-1}. More specifically, unlike Algorithm \ref{alg-1} in our proposed algorithm the nodes do not exchange local data $y_{k,i},\bH_{k,i},\bR_{k,i}$ with their neighbors in the incremental step and the algorithm solely relies on the transmission selected entries of $\bpsi_{k,i}$. So, in the proposed PDKF algorithm the incremental step \eqref{eq:2} changes to he following \emph{Adaptation Phase}:
\begin{align} \label{eq:p1}
					{\bR_{e,i}} & = {\bR_{k,i}} + {\bH_{k,i}}{\bP_{k,i|i}}\bH_{k,i}^* \nonumber \\
					{\bpsi _{k,i}} & = {\bpsi_{k,i}} + {\bP_{k,i|i}}\bH_{k,i}^*{\bR}_{e,i}^{-1}[{\by_{k,i}}-{\bH_{k,i}}{\bpsi_{k,i}}]   \\
          {\bP_{k,i}} & = {\bP_{k,i|i}}-{\bP_{k,i|i}}\bH_{k,i}^*\bR_{e,i}^{-1}{\bH_{k,i}}{\bP_{k,i|i}}
\nonumber
\end{align}
Selecting and scattering $L$ out of $M$, $0\leq L \leq M$, entries of the intermediate state estimate vector of each node $k$ at time instant $i$, make the realization of reducing internode communication possible. According to this scheme, the selection process can be implemented using a diagonal selection matrix, $\bT_{k,i}\in\R^{M \times M}$. Multiplication of $\bpsi_{k,i}$ by $\bT_{k,i}$ that has $L$ ones and $M-L$ zeros on its diagonal replaces its non-selected entries with zero. The positions of the ones on diagonal of $\bT_{k,i}$ determine the entries of node $k$ that are selected to diffuse at time $i$. Note that, the integer $L$ is fixed and pre-specified \cite{Arablouei2014a}.

\begin{figure} [t]
\centering 
\includegraphics [width=8.5cm]{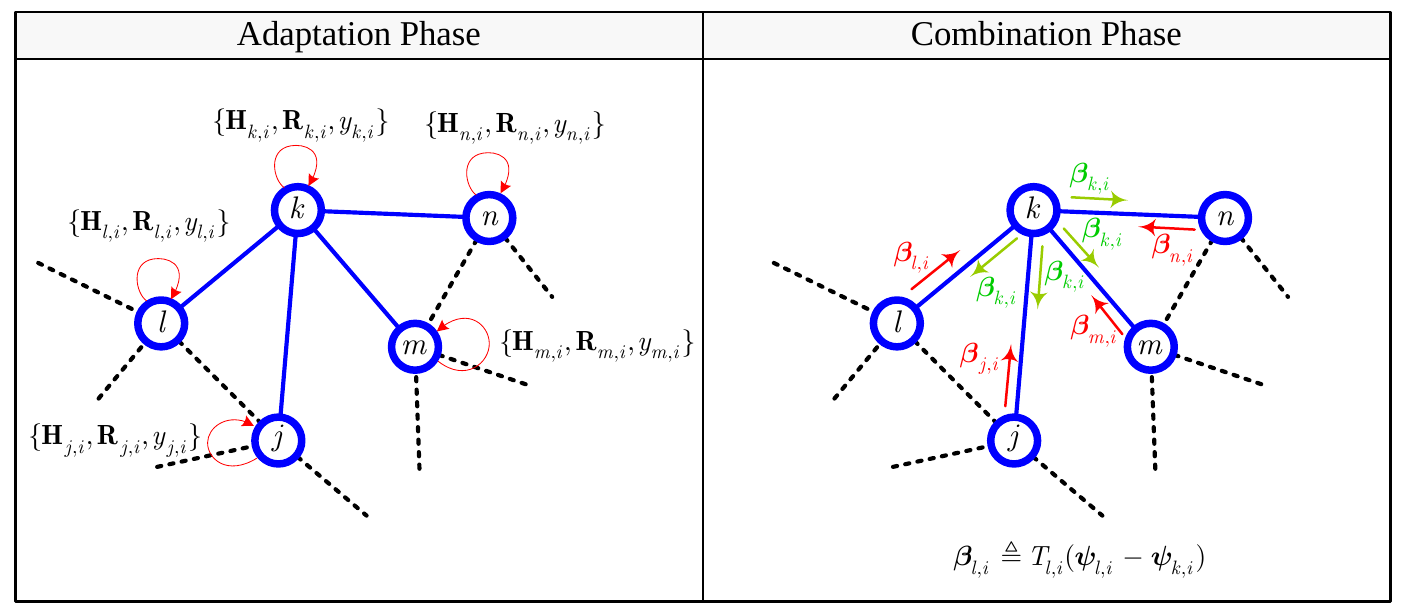} 
\centering \caption{Partial Diffusion Kalman filter update at node $k$.}
\label{fig:2}
\end{figure}

The most fundamental problem, we face with, hinges on ambiguities in non-diffused elements of the nodes in combination phase. When the intermediate estimates are partially transmitted, the non-communicated entries are not available to take part in this phase. However, each node requires all entries of the intermediate estimate vectors of its neighbors for combination. To avoid this ambiguity, the nodes can replace the entries of their own intermediate estimates instead of the ones from the neighbors that are not available. It would be useful to use the following equation (\emph{combination phase}) for aggregating the partially received intermediate estimates:
\begin{equation}
{\hat \bx_{k,i|i}} = {\bpsi _{k,i}} + \sum\limits_{l \in {{\cal N}_k}{\rm{\backslash \{ }}{\mathop{\rm k}\nolimits} {\rm{\} }}} {{c_{lk}}\bT_{l,i}{\rm{(}}{\bpsi _{l,i}} - {\bpsi _{k,i}}{\rm{)}}}
\label{eq:5}
\end{equation}
Therefore, we substitute the unavailable elements by their equivalent ones in each node's own intermediate estimate vector. Accordingly, our proposed PDKF algorithm employs \eqref{eq:p1} in the adaptation phase and \eqref{eq:5} in the combination phase. The proposed PDKF algorithm is described in Algorithm \ref{alg-2}. This process is also demonstrated schematically in Fig. \ref{fig:2}.
\begin{remark}
The computational complexity of the proposed PDKF algorithm is similar to the original DKF algorithm given by Algorithm \ref{alg-1}. To show this, note that \eqref{eq:5} can also be written as 
\begin{equation}
{\hat \bx_{k,i|i}} = { c_{kk}}{\bpsi _{ k,i}} + \sum\limits_{l \in {{\cal N}_k}{\rm{\backslash \{ }} k {\rm{\} }}} {c_{lk}[{{\bT}_{l,i}}{\bpsi _{{l,i}} + ({{\bI}_M}} - {\bT_{k,i}})} {\bpsi _{{k,i}}]}
\label{eq:6}
\end{equation}
Comparing \eqref{eq:3} and \eqref{eq:6} we can see that both expressions require $\left|{\cal N}_k\right|M$ multiplications and $\left(\left|{\cal N}_k\right|-1\right)M$ additions per iteration per node. 
\end{remark}

\subsection{Entry Selection Method}
To select $L$-subset of a set on $M$ elements containing exactly $L$ elements, we employ a similar approach proposed in \cite{Arablouei2014a}. Doing so, there exist two different schemes namely sequential and stochastic partial-diffusion. These methods are analogous to the selection processes in sequential and stochastic partial-update schemes \cite{Dogancay2008,Godavarti2005,Douglas1997}. In the sequential partial-diffusion the entry selection matrices, $\bT_{k,i}$ are diagonal:
\begin{equation}
\bT_{k,i} = 
\begin{bmatrix}
t_{1,i} & \cdots      &   0    \\
\vdots   &  \ddots & \vdots \\
0      & \cdots    & t_{M,i} 
\end{bmatrix}, \
t_{l,i}=
\begin{cases}
1  &  \text{if} \ l \in \mathcal{J}_{(i \mathrm{mod} \bar{\Omega})+1} \\
0  &   \text{otherwise}
\end{cases}
\label{eq:7}
\end{equation}
with $\bar \Omega = \left\lceil {M/L} \right\rceil $. The number of selection entries at each iteration is restricted by $L$. The coefficient subsets ${\cal J}_{i}$ are not unique as long as they meet the following requirements \cite{Dogancay2008}:
\begin{enumerate}
	\item Cardinality of $\mathcal{J}_i$ is between 1 and $L$;
	\item $\bigcup\nolimits_{\tau = 1}^{\bar \Omega} {{{\cal J}_\tau } = {\cal S}} $ where ${\cal S} = {\rm{\{ }}1,2, \ldots ,M{\rm{\} }}$;
	\item ${{\cal J}_\tau} \cap {{\cal J}_\upsilon} = \phi ,\forall \tau ,\upsilon \in {\rm{\{ 1,}} \ldots \bar \Omega\}$  and $\tau  \ne \upsilon$.
\end{enumerate}
The description of the entry selection matrices, $\bT_{k,i}$, in stochastic partial-diffusion is similar to that of sequential one. The only difference is as follows. At a given iteration $i$, for the sequential case, one of the set $\mathcal{J}_\tau$, $\tau  = 1, \ldots ,\bar \Omega$, set in advance, whereas for stochastic case, one of the sets $\mathcal{J}_\tau$ is sampled at random from $\{\mathcal{J}_1,\mathcal{J}_2,\cdots,\mathcal{J}_{\bar{\Omega}}\}$. We do this because the nodes need to know which entries of their neighbors' intermediate estimates have been transmitted at each iteration. These schemes bypass the need for any addressing procedure.

\begin{algorithm}[!t]
  \caption{Partial Diffusion Kalman filter}
  \begin{algorithmic}
	\State \textbf{Initialization:} $\hat\bx_{k,0|-1}=0$ and $\bP_{k,0|-1}=\boldsymbol{\Pi}_0$
	       \State For every time instant $i$, every node $k$ computes
			\State 
       \State \textbf{Step1:} Adaptation phase 
        \State ${\bpsi_{k,i}} \leftarrow {\hat\bx_{k,i|i-1}}$
		    \State ${\bP_{k,i}} \leftarrow {\bP_{k,i|i-1}}$  
				\For {$k=1,2,\cdots N$}
				\begin{align}
					{\bR_{e,i}} & = {\bR_{k,i}} + {\bH_{k,i}}{\bP_{k,i|i}}\bH_{k,i}^* \nonumber \\
					{\bpsi _{k,i}} & = {\bpsi_{k,i}} + {\bP_{k,i|i}}\bH_{k,i}^*{\bR}_{e,i}^{-1}[{\by_{k,i}}-{\bH_{k,i}}{\bpsi_{k,i}}] \nonumber  \\
          {\bP_{k,i}} & = {\bP_{k,i|i}}-{\bP_{k,i|i}}\bH_{k,i}^*\bR_{e,i}^{-1}{\bH_{k,i}}{\bP_{k,i|i}}
					\nonumber
					\end{align}
				\EndFor
        \State 
       \State \textbf{Step2:} Combination Phase
			\begin{align}
{\hat \bx_{k,i|i}} &\leftarrow {\bpsi _{k,i}} + \sum\limits_{l \in {{\cal N}_k}{\rm{\backslash \{ }}{\mathop{\rm k}\nolimits} {\rm{\} }}} {{c_{lk}}\bT_{l,i}{\rm{(}}{\bpsi _{l,i}} - {\bpsi _{k,i}}{\rm{)}}}
\nonumber \\
{\bP_{k,i|i}} & \leftarrow {\bP_{k,i}} \nonumber \\
{\hat\bx_{k,i+1|i}}&=\bF_{i}{\hat\bx_{k,i|i}} \nonumber \\
{\bP_{k,i+1|i}} &= \bF_{i}{\bP_{k,i|i}}\bF_{i}^*+\bG_{i}\bQ_{i}\bG_{i}^*   \nonumber 
\end{align}
  \end{algorithmic}
  \label{alg-2}
  \end{algorithm}

\section{Performance Analysis} \label{sec3}
\subsection{Network Update Equation}
In this section we present the mean, mean-square and convergence analysis of the proposed PDKF algorithm. We also derive a closed-form expression for MSD, which is frequently used as a steady-state metric in adaptive networks. For every node $k$, the MSD metric is defined as follows:
\begin{equation}
\MSD_{k,i}=\E{\|\bx_i-\hat{\bx}_{k,i|i}\|^2}
\label{eq:8}
\end{equation}
Let $\tilde{\bpsi}_{k,i}=\bx_i-\bpsi_{k,i}$ denote the estimation error at the end of the Adaptation phase. Then, it can be easily shown that the following expression holds
\begin{align}
\tilde{\bpsi}_{k,i}&=\tilde{\bx}_{k,i|i-1}-\bP_{k,i|i-1}\bH_{k,i}^*\bR_{e,i}^{-1}(\bH_{k,i}\tilde{\bx}_{k,i|i-1}+\bv_{k,i}) \nonumber \\
									 &=(\bI_{M}-\bP_{k,i|i-1}\bH_{k,i}^*\bR_{e,i}^{-1}\bH_{k,i})\tilde{\bx}_{k,i|i-1}\nonumber \\
									&-\bP_{k,i|i-1}\bH_{k,i}^*\bR_{e,i}^{-1} \bv_{k,i}
\label{eq:9}
\end{align}
Noting that $\bP_{k,i|i}\bH_{k,i}^*\bR_{e,i}^{-1}=\bP_{k,i|i-1}\bH_{k,i}^*\bR_{e,i}^{-1}$, the above equation can be rewritten as:  
\begin{equation}
\tilde{\bpsi}_{k,i}=(\bI_{M}-\bP_{k,i|i}\bS_{k,i})\tilde{\bx}_{k,i|i-1}-\bP_{k,i|i}\bH_{k,i}^*\bR_{e,i}^{-1} \bv_{k,i}
\label{eq:10}
\end{equation}
where $\bS_{k,i}=\bH_{k,i}^*\bR_{e,i}^{-1}\bH_{k,i}$. We also have 
\begin{equation}
\widetilde{\bx}_{k,i|i-1}=\bF_{i-1}\tilde{\bx}_{k,i-1|i-1}+\bG_{i-1}\bn_{i-1}
\label{eq:11}
\end{equation}
Substituting (13) into (12) gives
\begin{align}
\widetilde{\bpsi}_{k,i}&=(\bI_{M}-\bP_{k,i|i}\bS_{k,i})\bF_{i-1}\tilde{\bx}_{k,i-1|i-1} \nonumber \\
 & \hspace{0.2cm}  +(\bI_{M}-\bP_{k,i|i}\bS_{k,i})\bG_{i-1}\bn_{i-1}-\bP_{k,i|i}\bH_{k,i}^*\bR_{k,i}^{-1} \bv_{k,i}
\label{eq:12}
\end{align}
Now, to proceed let us define the augmented state-error vectors $\widetilde{\bmX}_{i|i}$ and $\widetilde{\bmPsi}_i$, measurement noise vectors $\bv_{i}$, and block-diagonal matrices $\bmH_i$, $\bmP_{i|i}$, $\bmS_{i}$ and $\bmB_{i}$ as follows:
\begin{align}
\widetilde{\bmX}_{i|i}&=\col{\widetilde{\bx}_{1,i|i},\widetilde{\bx}_{2,i|i},\cdots,\widetilde{\bx}_{N,i|i}} \nonumber \\
\widetilde{\bmPsi}_i&= \col{\widetilde{\bpsi}_{1,i},\widetilde{\bpsi}_{2,i},\cdots,\widetilde{\bpsi}_{N,i}} \nonumber \\
\bv_{i}&=\col{\bv_{1,i},\cdot,\bv_{N,i}}\nonumber \\
\bmH_i&= \Diag{\bH_{1,i},\bH_{2,i},\cdots,\bH_{N,i}} \nonumber \\
\bmP_{i|i}&= \Diag{\bP_{1,i|i},\bP_{2,i|i},\cdots,\bP_{N,i|i}} \nonumber \\
\bmS_i&= \Diag{\bS_{1,i},\bS_{2,i},\cdots,\bS_{N,i}} \nonumber \\
\bmB_i&=\begin{bmatrix}
\bB_{1,1,i} & \cdots      &   \bB_{1,N,i}    \\
\vdots   &  \ddots & \vdots \\
\bB_{N,1,i}      & \cdots    & \bB_{N,N,i} 
\end{bmatrix}.\nonumber
\end{align}
where
\begin{equation}
\bB_{p,q,i}=\begin{cases}
\bI_M-\sum_{l\in{\mathcal{N}_p}\backslash\left\{p\right\}}c_{lp}\bT_{l,i} & \text{if}\, p=q \\
c_{qp}\bT_{q,i} & \text{if}\, q\in{\mathcal{N}_p}\backslash\left\{p\right\} \\
\mathbf{O}_M & \text{otherwise}
\end{cases}\nonumber
\end{equation}
Using the above definitions, we now can express equations \eqref{eq:6} and \eqref{eq:12} in a global form as:
\begin{equation}
\widetilde{\bmX}_{i|i}=\bmB_{i}\widetilde{\bmPsi}_{i}
\label{eq:13}
\end{equation}
\begin{align}
\widetilde{\bmPsi}_{i}&=\left(\bI_{MN}-\bmP_{i|i}\bmS_{i}\right)[\left(\bI_{N}\otimes\bF_{i-1}\right)\widetilde{\bmX}_{i|i}\nonumber\\
& \hspace{0.4cm} +\left(\bI_{N}\otimes\bG_{i-1}\right)\left(\b1\otimes\bn_{i-1}\right)]-\bmP_{i|i}\bmH^{*}_i{\bR^{-1}_{i}}\bv_{i}
\label{eq:14}
\end{align}
where $\bR_{i}=\E{\bv_{i}\bv^{*}_{i}}$ is a block-diagonal matrix. Note that equation \eqref{eq:13} describes the evolution of entire network. Note further that \eqref{eq:14} can be rewritten in a more compact form as:
\begin{equation}
\widetilde{\bmPsi}_{i}=\bmF_{i}\widetilde{\bmX}_{i-1|i-1}+\bmG_{i}\left(\b1\otimes\bn_{i-1}\right)-\bmD_{i}\bv_{i}
\label{eq:15}
\end{equation}
where
\begin{align}
\bmF_{i}&=\left(\bI_{MN}-\bmP_{i|i}\bmS_{i}\right)\left(\bI_{N}\otimes\bF_{i-1}\right)\nonumber\\
\bmG_{i}&=\left(\bI_{MN}-\bmP_{i|i}\bmS_{i}\right)\left(\bI_{N}\otimes\bG_{i-1}\right)\nonumber\\
\bmD_{i}&=\bmP_{i|i}\bmH^{*}_i{\bR^{-1}_{i}}\nonumber
\end{align}

By substituting \eqref{eq:15} into \eqref{eq:13} the update equation for the network state-error vector becomes
\begin{equation}
\widetilde{\bmX}_{i|i}=\bmB_{i}\bmF_{i}\widetilde{\bmX}_{i-1|i-1}+\bmB{i}\bmG_{i}\left(\b1\otimes\bn_{i-1}\right)-\bmB_{i}\bmD_{i}\bv_{i}
\label{eq:16}
\end{equation}
\subsection{Assumptions}
As it is common to impose statistical assumptions on the regression and noise data to make the analysis tractable, in the next Section, we will consider the following assumptions in our analysis.
\begin{assumption}\label{asp:2}\
\begin{enumerate}[(i)]
\item The matrices in the model described \eqref{eq:1} and \eqref{eq:1b} are time-invariant and the noise is considered stationary, i.e., the matrices $\bF$, $\bG$, $\bH$, $\bR$ and $\bQ$ do not depend on time $i$. Furthermore, we assume that matrix $\bF$ is stable, the necessary and sufficient condition for stability of $\bF$ is that all its eigenvalues lie strictly inside the unite circle. 
\item Let define the \emph{local} Kalman filter as a A Kalman filter that uses data from its neighborhood. Then, we assume that local Kalman filter converges for every neighborhood, i.e., $\lim_{i\to \infty}\bP_{k,i|i-1}\triangleq {\bP}^{-}_{k}$ and $\lim_{i\to \infty}\bP_{k,i|i}\triangleq\bP_{k}$, $k=\left\{1,\cdots,N\right\}$, called ergodicity assumption.
\end{enumerate}
\end{assumption}
Under these assumptions, the matrices $\bmF_{i}$, $\bmG_{i}$ and $\bmD_{i}$ also converge in steady-state, and their corresponding steady-state values are given by
\begin{align}
\bmP&\triangleq\lim_{i\to \infty}\bmP_{i|i}=\Diag{\bP_{1},\ldots,\bP_{N}}\nonumber\\
\bmP^{-}&\triangleq\lim_{i\to \infty}\bmP_{i|i-1}=\Diag{\bP^{-}_{1},\ldots,\bP^{-}_{N}}\nonumber\\
\bmF&\triangleq\lim_{i\to \infty}\bmF_{i}=\left(\bI_{MN}-\bmP\bmS\right)\left(\bI_{N}\otimes\bF\right)\nonumber\\
\bmG&\triangleq\lim_{i\to \infty}\bmG_{i}=\left(\bI_{MN}-\bmP\bmS\right)\left(\bI_{N}\otimes\bG\right)\nonumber\\
\bmD&\triangleq\lim_{i\to \infty}\bmD_{i}=\bmP\bmH^{*}{\bR^{-1}}\nonumber
\end{align}
where $\bmS$ and $\bmH$ are used instead of $\bmS_{i}$ and $\bmH_{i}$ since these matrices are now time-invariant.

\subsection{Mean Performance}
Since $\hat{\bx}_{k,0|-1}$ and $\E{\bx_{0}}=0$, we have $\E{\hat{\bx}_{k,0|-1}}$ for all $k$. Moreover, all entries of $\bQ$ are real non-negative and each row summing to $(\bQ \mathds{1}=\mathds{1})$ which means that $\bQ$ is a right-stochastic matrix. Therefore, we have 
\begin{align}
\lim_{i\to \infty}\E{\widetilde{\bmX}_{i|i}}&=\bmO_{MN}\nonumber
\end{align}
where $\bmO_{MN}$ denotes the $MN\times 1$  zero vector. This means the PDKF algorithm is convergent in the mean sense and asymptotically unbiased.

\subsection{Mean-Square Performance}
Taking the squared weighted Euclidean norm of both sides of \eqref{eq:16} and applying the expectation operator together with using \emph{Assumption} \ref{asp:2} yield the following weighted variance relation:
\begin{align}
\E{\left\|\widetilde{\bmX}_{i|i}\right\|^{2}_{\Sigma}}&=\E{\left\|\widetilde{\bmX}_{i-1|i-1}\right\|^{2}_{\Gamma}}+\Ee{\rm{[}}\left(\b1\otimes\bn_{i-1}\right)^{*}\bmG^{*}_{i}\bmB^{*}_{i}\bmSig\nonumber\\
&\bmB_{i}\bmG_{i}\left(\b1\otimes\bn_{i-1}\right){\rm{]}}+\E{\bv^{*}_{i}\bmD^{*}_{i}\bmB^{*}_{i}\bmSig\bmB_{i}\bmD_{i}\bv_{i}}
\label{eq:17}
\end{align}
\begin{align}
\bmGa&=\bmB^{*}_{i}\bmF^{*}_{i}\bmSig\bmF_{i}\bmB_{i}
\label{eq:18}
\end{align}
where $\bmSig$ is an arbitrary symmetric nonnegative-definite matrix. Since $\widetilde{\bmX}_{i-1|i-1}$ is independent of $\bmGa$ we have,
\begin{equation}
\E{\left\|\widetilde{\bmX}_{i-1|i-1}\right\|^{2}_{\Gamma}}=\E{\left\|\widetilde{\bmX}_{i-1|i-1}\right\|^{2}_{\E{\Gamma}}}
\label{eq:19}
\end{equation}
Define
\begin{equation}
	\gamma=\vect{\E{\bmGa}}
	\label{eq:20}
\end{equation}
	
	\begin{equation}
\boldsymbol{\sigma}=\vect{\bmSig}
\label{eq:21}
	\end{equation}
where $\vect{.}$ denotes a linear transformation which converts the matrix into a column vector by stacking all columns of its matrix argument. The transpose of a vectorized matrix are also denoted by $\mathop{\rm {vec}}^{T}\left\{.\right\}$. Using \eqref{eq:19}-\eqref{eq:21}, we can alter \eqref{eq:17} to 
\begin{align}
\E{\left\|\widetilde{\bmX}_{i|i}\right\|^{2}_{\sigma}}&=\E{\left\|\widetilde{\bmX}_{i-1|i-1}\right\|^{2}_{\gamma}} \nonumber\\
& \hspace{0.5cm} +\E{\left(\b1\otimes\bn_{i-1}\right)^{*}\bmG^{*}_{i}\bmB^{*}_{i}\bmSig
 \bmB_{i}\bmG_{i}\left(\b1\otimes\bn_{i-1}\right){\rm{]}}} \nonumber\\
& \hspace{0.5cm} +\E{\bv^{*}_{i}\bmD^{*}_{i}\bmB^{*}_{i}\bmSig\bmB_{i}\bmD_{i}\bv_{i}}
\label{eq:22}
\end{align}
where $\E{\left\|\widetilde{\bmX}_{i|i}\right\|^{2}_{\sigma}}$ and $\E{\left\|\widetilde{\bmX}_{i-1|i-1}\right\|^{2}_{\gamma}}$ are the same quantities as $\E{\left\|\widetilde{\bmX}_{i|i}\right\|^{2}_{\Sigma}}$ and $\E{\left\|\widetilde{\bmX}_{i-1|i-1}\right\|^{2}_{\Gamma}}$ , respectively.

Using \eqref{eq:18}, \eqref{eq:20}, and \eqref{eq:21}, the commutative property of the expectation and vectorization operations, and the relationship between the Kronecker product and the vectorization operator \cite{Dogancay2008}, i.e.,
\begin{align}
\vect{\mathbf{ABC}}=\left(\mathbf{C}^{T}\otimes \mathbf{A}\right)\vect{\mathbf{B}}\nonumber
\end{align}
We can verify that
\begin{equation}
\boldsymbol{\gamma}=\bmfB\left(\bmF^{T}_{i}\otimes\bmF^{*}_{i}\right)\boldsymbol{\sigma}
\label{eq:23}
\end{equation}
where
\begin{align}
	\bmfB=\E{\bmB^{T}_{i}\otimes\bmB^{*}_{i}}\nonumber
\end{align}
Using the following property from linear algebra \cite{Arablouei2014}
\begin{align}
\mathrm{tr}\left(\mathbf{A}^{T}\mathbf{B}\right)=\mathrm{vec}^{T}\left\{\mathbf{B}\right\}\vect{\mathbf{A}}\nonumber
\end{align}
and the symmetry of $\bmSig$ , we have
\begin{IEEEeqnarray}{l}
\E{\left(\b1\otimes\bn_{i-1}\right)^{*}\bmG^{*}_{i}\bmB^{*}_{i}\bmSig\bmB_{i}\bmG_{i}\left(\b1\otimes\bn_{i-1}\right)}\nonumber\\
=\E{\mathrm{tr}\left\{\bmSig\bmB_{i}\bmG_{i}\left(\b1\otimes\bn_{i-1}\right)\left(\b1\otimes\bn_{i-1}\right)^{*}\bmG^{*}_{i}\bmB^{*}_{i}\right\}}\nonumber\\
=\mathrm{vec}^{T}\left\{\E{\left\{\bmSig\bmB_{i}\bmG_{i}\left(\b1\otimes\bn_{i-1}\right)\left(\b1\otimes\bn_{i-1}\right)^{*}\bmG^{*}_{i}\bmB^{T}_{i}\right\}}\right\}\nonumber\\
=\left(\bmfB^{T}\mathrm{vec}\left\{\E{\bmG_{i}\left(\b1\otimes\bn_{i-1}\right)\left(\b1\otimes\bn_{i-1}\right)^{*}\bmG^{*}_{i}}\right\}\right)^{T}\boldsymbol{\sigma}\nonumber\\
=\mathrm{vec}^{T}\left\{\bmL\right\}\bmfB\boldsymbol{\sigma}
\label{eq:24}
\end{IEEEeqnarray}
where
\begin{align}
\bmL=\bmG_{i}\left(\b1\b1^{T}\otimes\bmQ_{i-1}\right)\bmG^{*}_{i}\nonumber
\end{align}
The last term of \eqref{eq:17} can be written as
\begin{align}
\E{\bv^{*}_{i}\bmD^{*}_{i}\bmB^{*}_{i}\bmSig\bmB_{i}\bmD_{i}\bv_{i}}&=\E{\mathrm{tr}\left\{\bmSig\bmB_{i}\bmD_{i}\bv_{i}\bv^{*}_{i}\bmD^{*}_{i}\bmB^{*}_{i}}\right\}\nonumber\\
&=\mathrm{vec}^{T}\left\{\bmB_{i}\bmD_{i}\bv_{i}\bv^{*}_{i}\bmD^{*}_{i}\bmB^{T}_{i}\right\}\boldsymbol{\sigma}\nonumber\\
&=\left(\bmfB^{T}\vect{\E{\bmD_{i}\bv_{i}\bv^{*}_{i}\bmD^{*}_{i}}}\right)^{T}\boldsymbol{\sigma}\nonumber\\
&=\mathrm{vec}^{T}\left\{\bmK\right\}\bmfB\boldsymbol{\sigma}
\label{eq:25}
\end{align}
where
\begin{align}
\bmK=\bmD_{i}\bR_{i}\bmD^{*}_{i}\nonumber
\end{align}
Substitution of \eqref{eq:25}, \eqref{eq:24} and \eqref{eq:23} into \eqref{eq:22} gives
\begin{align}\label{eq:26}
\E{\left\|\widetilde{\bmX}_{i|i}\right\|^{2}_{\boldsymbol{\sigma}}}&=\E{\left\|\widetilde{\bmX}_{i-1|i-1}\right\|^{2}_{\bmfB\left(\bmF^{T}_{i}\otimes\bmF^{*}_{i}\right)\boldsymbol{\sigma}}}\nonumber\\
&\hspace{0.7cm}+\mathrm{vec}^{T}\left\{\bmL\right\}\bmfB\boldsymbol{\sigma}+\mathrm{vec}^{T}\left\{\bmK\right\}\bmfB\boldsymbol{\sigma}
\end{align}
Taking the limitation as $i\to \infty$, we obtain from \eqref{eq:26} that:
\begin{equation}\label{eq:27}
\lim_{i\to \infty}\E{\left\|\widetilde{\bmX}_{i|i}
\right\|^{2}_{\left(\bI_{M^{2}N^{2}}-\bmfB \left(\bmF^{T}_{i}\otimes\bmF^{*}_{i}\right)\right)\boldsymbol{\sigma}}}=\mathrm{vec}\left\{\bmK+\bmL\right\}\bmfB\boldsymbol{\sigma}
\end{equation}
Note that since $\bmF$ is stable, the matrix $\bI_{M}-\bmfB\left(\bmF^{T}_{i}\otimes\bmF^{*}_{i}\right)$ is non-singular.
Finally, the average steady-state MSD across the network is
\begin{align}
\mathrm{MSD}^{\mathrm{network}}&=\frac{1}{N}\mathrm{vec}^{T}\left\{\bmK+\bmL\right\}\bmfB \times \nonumber\\
 &\hspace{0.7cm}\left(\bI_{M^{2}N^{2}}-\bmfB \left(\bmF^{T}_{i}\otimes\bmF^{*}_{i}\right)\right)^{-1} \mathrm{vec}\left\{\bI_{MN}\right\}
\label{eq:28}
\end{align}
We can summarize the results of Section III in the following Lemma.
\begin{lem}
Under Assumptions \ref{asp:1} and \ref{asp:2}, the partial diffusion KF algorithm given by Algorithm \ref{alg-2} is convergent in mean and asymptotically unbiased, and the steady-state network MSD is given by \eqref{eq:28}. 
\end{lem}

\section{Simulations}
We now illustrate the simulation results for partial diffusion KF algorithm \eqref{eq:8}, and draw a comparison between its performance and the theoretical results of section III. doing so, we present a simulation example in Figs. (4) and (5). The measurements were generated according to model \eqref{eq:1}. The network is randomly generated and has a total of $N=10$ nodes where each node is, on average, linked with two other nodes. The size of the unknown parameter of the system is $M=8$. The state of the system is unknown 2-dimensional vector location of an object, i.e. $\left(x y\right)$, where $x$ and $y$ are first and second entries, respectively.  The state-space model matrices in \eqref{eq:1} are:
\begin{equation}
\bF = 
\begin{bmatrix}
1 & 0 & 0.1 & 0 \\
0 & 1 & 0 & 0.1 \\
0 & 0 & 1 & 0 \\
0 & 0 & 0 & 1
\end{bmatrix}, 
\bG=0.625\bI_{4}, 
\bQ=0.001\bI_{4}. \nonumber
\end{equation}
We assume that every node measures the position of the unknown object in either the two horizontal dimensions, or a combination of one horizontal and one vertical dimension. Thus, individual nodes do not have direct measurements of the position in the three dimensions. Therefore, we have
\begin{equation}
\bH_{k,i} = 
\begin{bmatrix}
0 & 1 & 0 & 0 \\
0 & 0 & 1 & 0 \\
0 & 0 & 0 & 0 \\
\end{bmatrix},\nonumber
\bH_{k,i} =
\begin{bmatrix}
0 & 1 & 0 & 0 \\
0 & 0 & 0 & 0 \\
0 & 0 & 0 & 1 \\
\end{bmatrix}.\nonumber
\end{equation}
at random, but with the requirement that every neighborhood should have nodes with both types of matrices to guarantee  that the convergence of local Kalman filter follows \emph{Assumption} \eqref{asp:2}. Finally, the measurement noise covariance matrix for agent $k$ is, $\bR_{k,i}=\sigma^{2}_{k,i}\bI_{3}$, where the noise variance $\sigma^{2}_{k,i}$ across agents  is selected randomly in the range [0 0.5]. The experimental results are achieved by ensemble averaging over 200 independent runs. The Steady-state values are also assessed by taking average over 1000 iterations. We also use uniform combination rule \cite{Cattivelli2010a} in the combination phase and initialize the estimates to zero vectors. The state noise covariance matrix traces and observation noise variances for all nodes are illustrated in Fig. 3. In Figs. 4 and 5, we plot the average network MSD curves of the PDKF algorithm, versus $i$, using both sequential and stochastic partial-diffusion schemes for different numbers of entries transmitted at each time update, $L$. Experimental and theoretical steady-state MSDs of all nodes for different numbers of entries transmitted at each iteration, $L$, are illustrated in Figs. 6 and 7. 
\begin{figure} [t]
\centering 
\includegraphics [width=9.5cm]{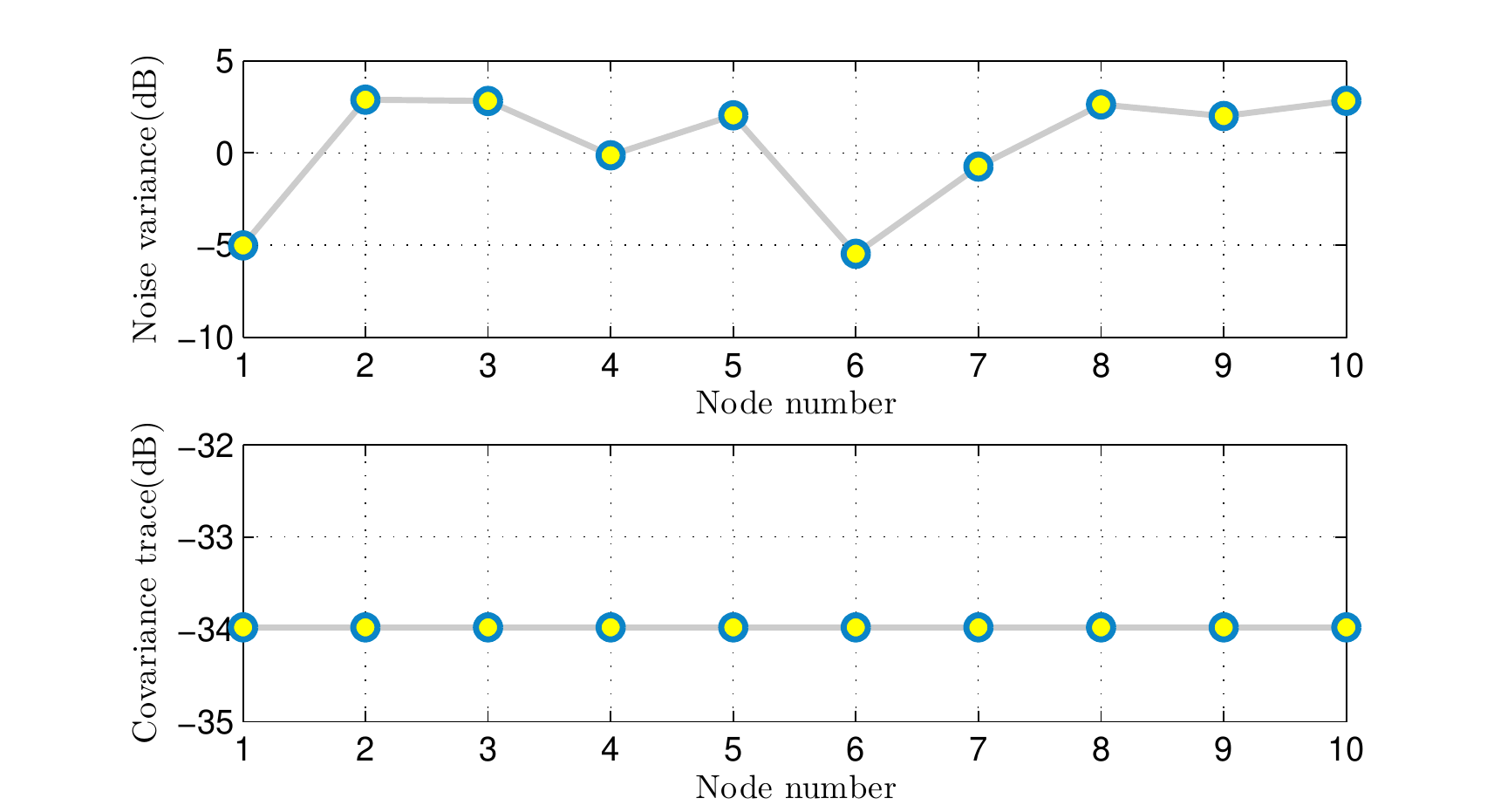} 
\centering \caption{Observation noise variance and covariance matrix trace of state noise at each node.}
\label{fig:3}
\end{figure}
\begin{figure} [t]
\centering 
\includegraphics [width=9.5cm]{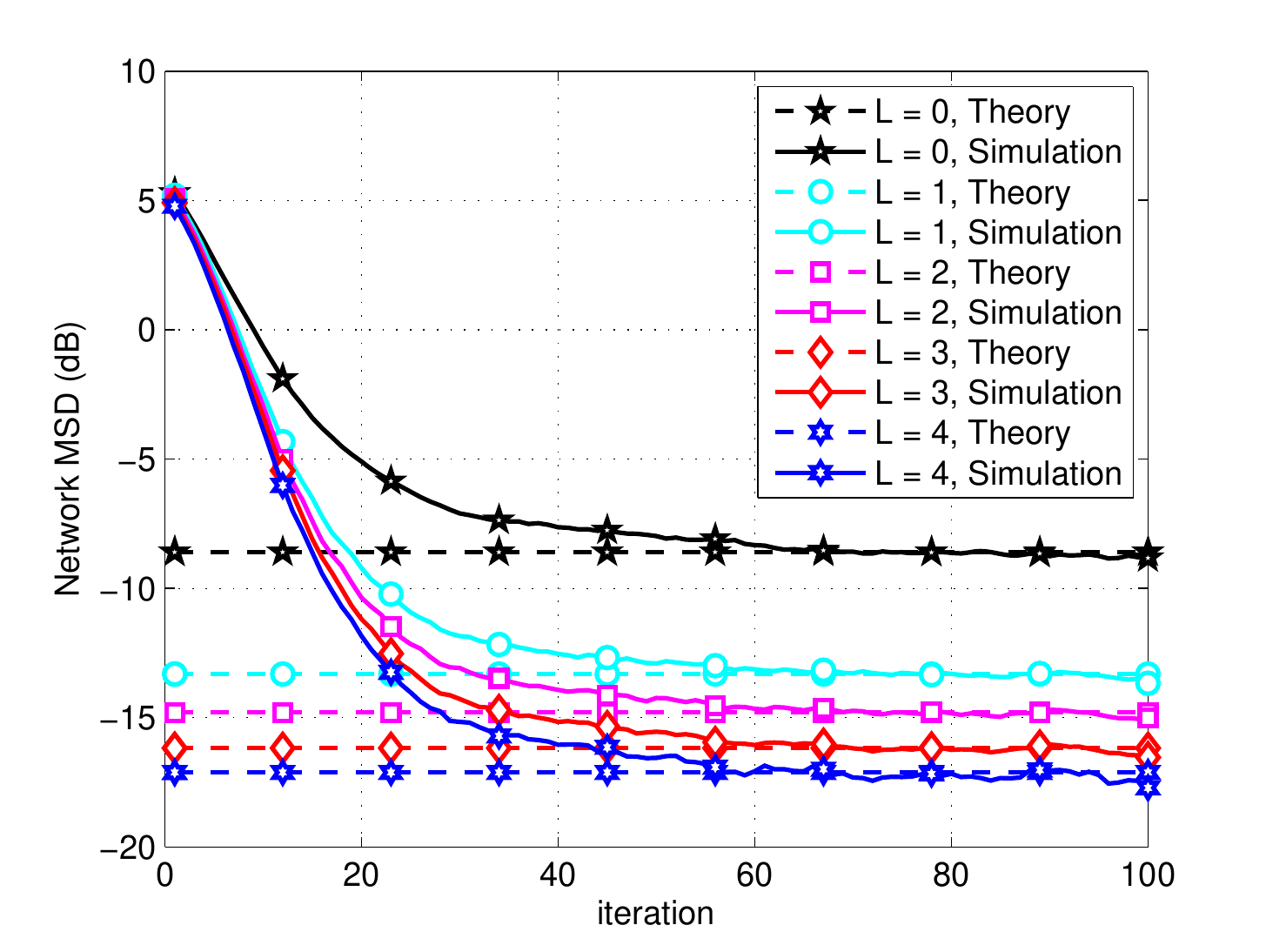} 
\centering \caption{Network MSD curves of PDKF algorithm using sequential partial-diffusion scheme for different numbers of entries communicated at each iteration.}
\label{fig:4}
\end{figure}
\begin{figure} [t]
\centering 
\includegraphics [width=9.5cm]{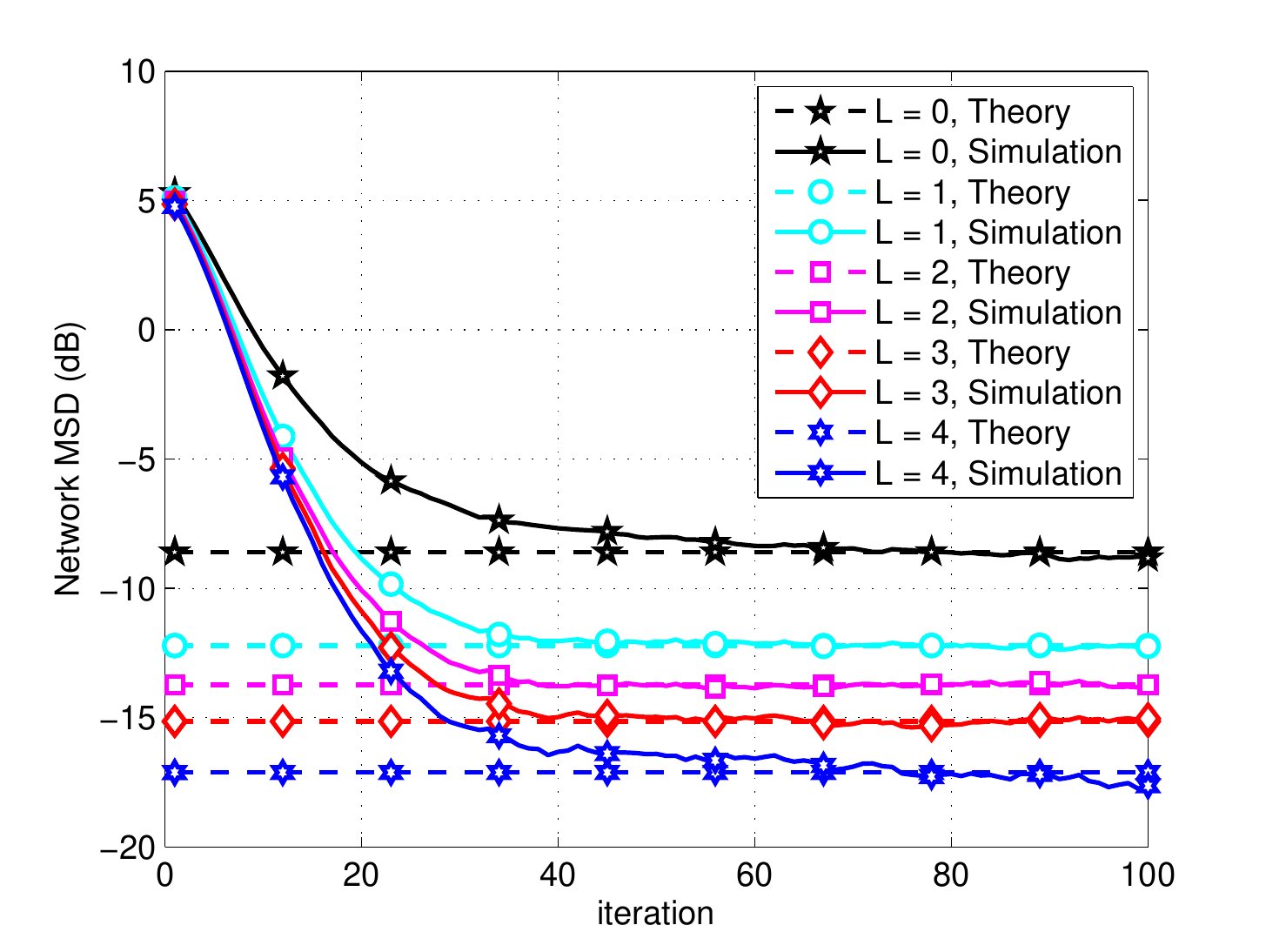} 
\centering \caption{Network MSD curves of PDKF algorithm using stochastic partial-diffusion scheme for different numbers of entries communicated at each iteration.}
\label{fig:5}
\end{figure}
\begin{figure} [th]
\centering 
\includegraphics [width=9.5cm]{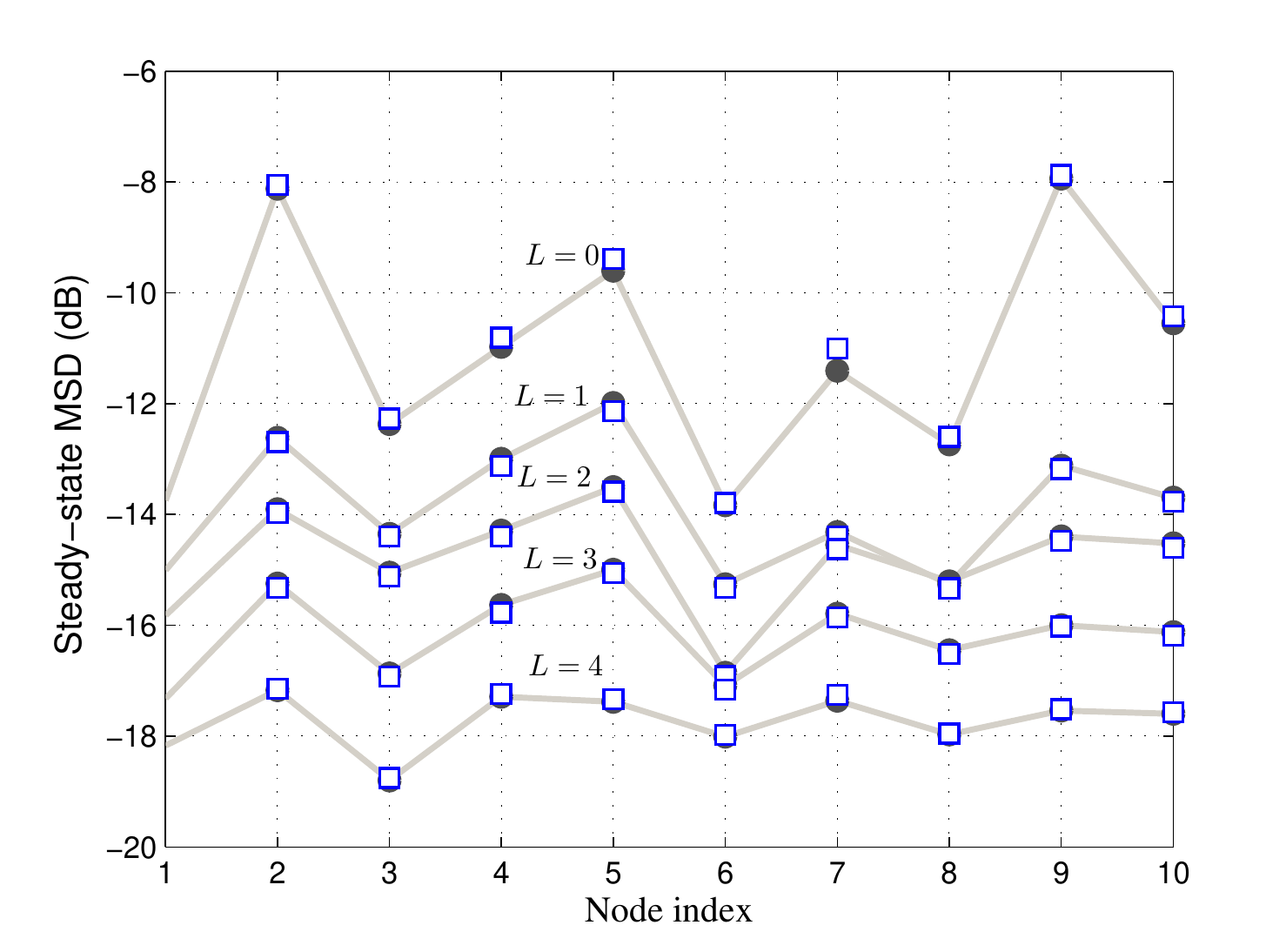} 
\centering \caption{Theoretical and experimental steady-state MSDs at each node for different values of $L$ using stochastic partial-diffusion scheme in the PDKF algorithm.  Note that the 
	points denoted by symbol $``\square"$ represent the results obtained from theory.}
\label{fig:6}
\end{figure}
\begin{figure} [th]
\centering 
\includegraphics [width=9.5cm]{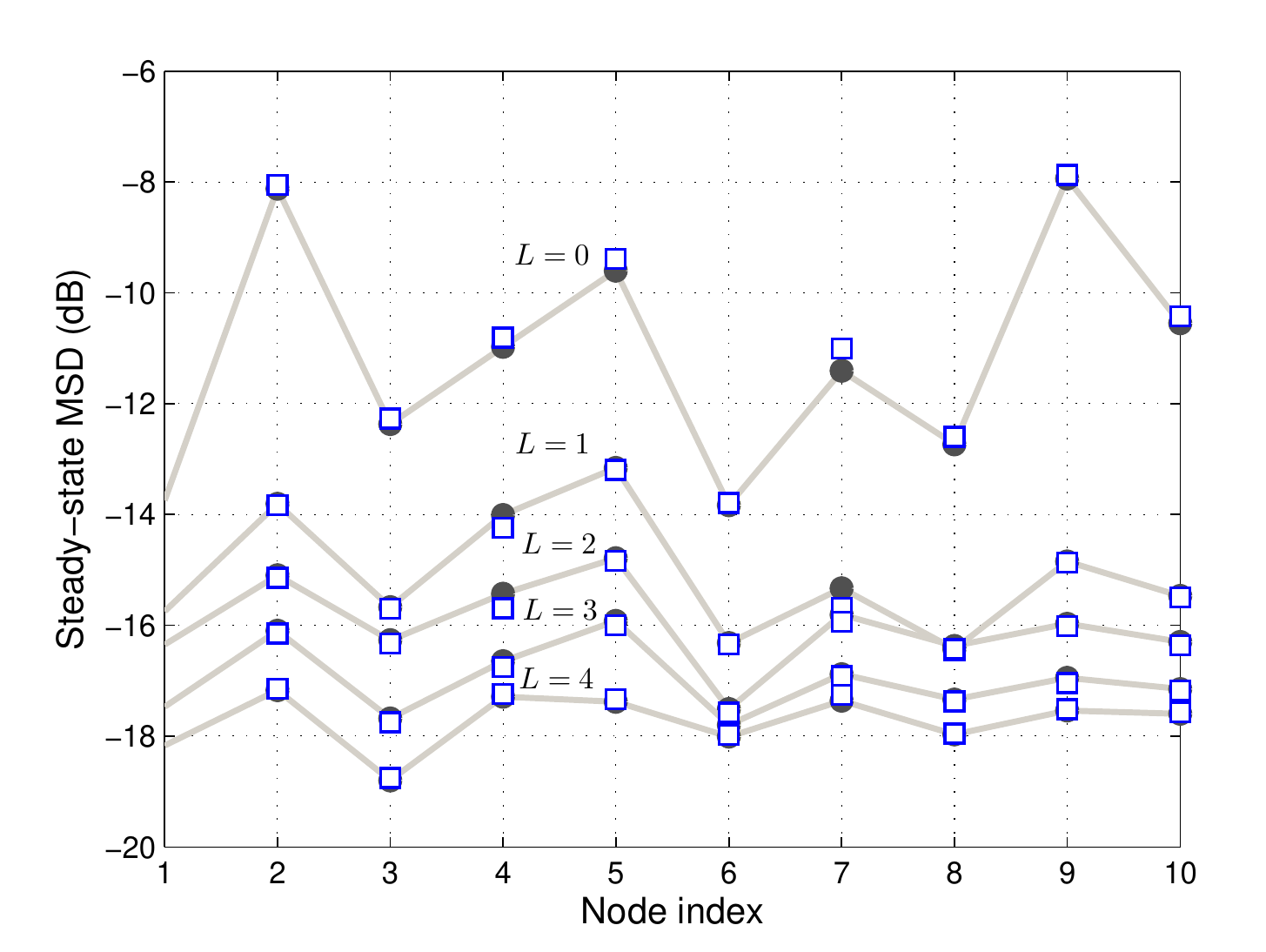} 
\centering \caption{Theoretical and experimental steady-state MSDs at each node for different values of $L$ using sequential partial-diffusion scheme in the PDKF algorithm. Note that the 
	points denoted by symbol $``\square"$ represent the results obtained from theory.}
\label{fig:6}
\end{figure}

From the results above, the following observations can be made:
\begin{itemize}
	\item The PDKF algorithm improves the usage of network transmission resources and provides a trade-off between communication density and estimation performance.
	\item The performance deterioration inspired by partial diffusion in comparison with savings attained in communication is negligible.
	\item A good match between the theoretical expression for the PDKF algorithm, obtained for the network MSD, and the experimental results are accomplished for different values of $L$.
\end{itemize}

\section{Conclusion}
In this correspondence, we examined partial-diffusion Kalman filtering (PDKF) algorithm for distributed state estimation. The PDKF enables reduced internode communications by allowing each node to transmit only a subset of intermediate state estimates to its close neighbors at every iteration. This fact directly leads to savings in bandwidth usage as well as power consumption of communication resources. To select the entries, communicating at each iteration, the nodes employ two different protocols namely stochastic and sequential partial-diffusion schemes. Consequently, using the PDKF algorithm, the required communications between the nodes are decreased in contrast to the case that all entries of the intermediate estimates are continuously transmitted. We analyzed the convergence of the algorithms and provided steady-state mean and mean-square analysis, showing a good agreement with the simulation results. Theoretical analysis and numerical simulations provided valuable insights into the performance of PDKF algorithm and illustrated that it offers a trade-off between communication density and estimation performance.

\balance

\end{document}